\documentclass[superscriptaddress]{revtex4}  
\RequirePackage{fix-cm}
\usepackage{graphicx}
\usepackage{color}
\usepackage{amsmath}
\usepackage{multirow}
\begin{document}

\title{Modelling of the dielectric properties of trabecular bone samples at microwave frequency}


\author{Ramiro M. Irastorza}
\email{rirastorza@iflysib.unlp.edu.ar}
\affiliation{Instituto de F\'{i}sica de L\'{i}quidos y Sistemas Biol\'{o}gicos (CONICET) Calle 59 No 789, B1900BTE La Plata, Argentina.}
\author{Eugenia Blangino}
\affiliation{Gabinete de Biomec\'{a}nica, Dto. de Ing. Mec\'{a}nica, Facultad de Ingenier\'{i}a, UBA.}
\author{Carlos M. Carlevaro}
\affiliation{Instituto de F\'{i}sica de L\'{i}quidos y Sistemas Biol\'{o}gicos (CONICET) Calle 59 No 789, B1900BTE La Plata, Argentina.}
\affiliation{Universidad Tecnol\'{o}gica Nacional - FRBA, UDB F\'{i}sica, Mozart No 2300, C1407IVT Buenos Aires, Agentina.}
\author{Fernando Vericat}
\affiliation{Instituto de F\'{i}sica de L\'{i}quidos y Sistemas Biol\'{o}gicos (CONICET) Calle 59 No 789, B1900BTE La Plata, Argentina.}
\affiliation{Grupo de Aplicaciones Matem\'{a}ticas y Estad\'{i}sticas de la Facultad de Ingenier\'{i}a UNLP, Argentina.}

\date{Received: date / Accepted: date}

\begin{abstract}
In this paper the dielectric properties of human trabecular bone are evaluated under physiological condition in the microwave range. Assuming a two components medium, simulation and experimental data are presented and discussed. A special experimental setup is developed in order to deal with inhomogeneous samples. Simulation data are obtained using finite difference time domain from a realistic sample. The bone mineral density of the samples are also measured. The simulation and experimental results of the present study suggest that there is a negative relation between bone volume fraction (BV/TV) and permittivity (conductivity): the higher the BV/TV the lower the permittivity (conductivity). This is in agreement with the recently published in vivo data. \\
\textit{Keywords}: Bone dielectric properties, Microwave tomography, Finite difference time domain.
\end{abstract}

\maketitle

\section{Introduction}
\label{INTRO}
One of the reason of the active search of alternative clinical methods to evaluate bone health is the undesirable ionizing radiation condition of the gold standard: bone mineral density (BMD) obtained via Dual X-ray Absorption (DXA) \cite{johnell2005}. Low and radio frequency dielectric properties evaluation appeared as non ionizing and low cost potential tools for non invasive body sensing. Currently, microwave tomography is an established field in biomedical imaging \cite{meaney2013microwave,meaney2013integration,persson2013microwave,kwon2012microwave,hagness2012guest,semenov2005microwave,liu2002active,mojabi2011novel}. Particularly for bone, the first clinical microwave tomographic images of the calcaneus were recently published \cite{meaney2012}. It means that with the current sensibility of the method, the dielectric properties of trabecular bone can be measured. \\
BMD is only one factor associated with the risk of fracture \cite{reginster2006}, other indicators are also used \cite{kanis2008}. Among them, micro-structure and its relationship with biomechanic response is one of the most studied characteristic of bone. Images are acquired by micro-Computed Tomography ($\mu$-CT) and the structure parameters are evaluated, i.e. bone volume fraction (BV/TV), degree of anisotropy (DA), trabecular thickness (ThTb), etc. It should be noted that it is also an ionizing technology. Some studies have indicated the degree of porosity as the only macroscopic difference between cortical and trabecular bone tissues \cite{tassani2011}. If this assumption is valid then the predictions of mechanical properties of bone tissue become a function of only bone quantity and structure, and hence, not a function of tissue mineral density. One interesting feature of the dielectric properties of bone is that information of micro-structure can be obtained by using permittivity models \cite{
irastorza2013,cherkaev2011,bonifasi2009,sierpowska2006,demercato1991}.\\
The degree of mineralisation and micro-structure of bone and its relation with dielectric properties was previously studied \cite{sierpowska2006,meaney2012bone,irastorza2011,ivancich1992,sierpowska2007}. The \textit{in vitro} radio frequency and microwave measurements of porcine trabecular samples \cite{meaney2012bone} (500-2500 MHz) are in agreement with the results presented in \cite{irastorza2011} (80-1000 MHz) and \cite{ivancich1992} (up to 1000 MHz) for bovine trabecular and cortical bone. These papers intend to explain the relation between mineralisation and dielectric properties by chemically modifying the mineral content of the samples. Both works show negative correlation between mineral density and relative permittivity and conductivity. The work of Meaney et al. \cite{meaney2012} compared ultrasound densitometry measures (among other parameters) to the microwave images \textit{in vivo} (for a frequency range of 500 MHz to 3 GHz). They have measured the human heels of two patients. Both patients had a noticeable low bone density in their left heels. The authors obtained a clear difference between permittivity and conductivity and mineral density (consistent with Houndsfield 
unities): the higher the bone density the lower the permittivity and conductivity. On the contrary, Sierpowska et al. \cite{sierpowska2006} found a positive correlation between BV/TV and relative permittivity ($R =$ 0.68), and a negative one with conductivity ($R = -$ 0.59) at 1.2 MHz. The same group found a similar behaviour between BMD and permittivity and conductivity ($R =$ 0.67 and $R = -$ 0.50, respectively) at 1.2 MHz as well. These examples show that the relationship between dielectric models, BV/TV and BMD is still an open problem. Until now, it seems that for frequencies higher than 700 MHz the dielectric parameters of animal bone samples would have a negative correlation with mineral degree and with bone volume. Remarkably, the chemical alteration of mineral content make these results relatively comparable with the \textit{in vivo} case.\\
In this paper we study the relationship between BV/TV and the dielectric properties of human trabecular samples \textit{in vitro} at physiological state. Samples were taken from different patients and, as 
a result, they have different degree of mineralisation. Samples are minimally manipulated without any chemical treatment, therefore we observe only the natural biological variability of them.\\
Simulation of the electric and the dielectric properties of bone and other tissues \cite{bonifasi2009,katz2006} are usually performed using the classical reference data of Gabriel et al. \cite{gabriel1996a,gabriel1996b,gabriel1996c}. For example, in the simulation work of Bonifasi-Lista and Cherkaev \cite{bonifasi2009}, the trabecular bone porosity is recovered from effective dielectric properties. The authors assume that the trabecular bone can be dielectrically interpreted as a two components heterogeneous medium: trabecular bone matrix with cavities filled with bone marrow.\\
The purpose of this study is to investigate the relationship between BV/TV and the dielectric properties at high frequencyof cancellous bone from the human femur head of elderly donors (mean age 80.7, range 74-94 years). The dielectric measurements are performed using open-ended coaxial lines (OECL) with a previously designed protocol developed for inhomogeneous samples \cite{irastorza2011}. The selected frequency range is between 100 MHz and 1300 MHz, near to the usually chosen range of the current \textit{in vivo} microwave tomography equipments \cite{meaney2012}.\\
In addition, we intend to give a step towards the modelling of the dielectric properties of trabecular bone tissue. We hypothesized that the use of dielectric properties would permit to obtain micro-structure information from \textit{in vivo} microwave tomographic images. Such knowledge would provide additional information to predict the mechanical properties. In this regard, simulations of effective dielectric properties of trabecular bone and the correlation with BV/TV are presented.

\section{Materials and Methods}
\label{MATMETH}
\subsection{Sample preparation}
\label{SAMPLE}
Six femoral heads were obtained from patients (mean age 80.7, range 74-96 years) undergoing total hip replacement. The live donors were 5 females and 1 male and all of them signed the corresponding informed consent. The surgical extraction was non traumatic to preserve the integrity of the bone tissue. The femoral heads were frozen at -20 $^{o}$C immediately after the extraction. Each femoral head was cut with custom tools extracting a cylindrical sample of 11 mm in diameter and always greater than 15 mm in length. The cylinders were placed in phosphate buffered saline (PBS) and frozen at -20 $^{o}$C, until tests were performed.  
 
\subsection{Bone mineral content and bone volume fraction}
\label{BMD}
The bone mineral content (BMC in grams) was measured in each cylinder by DXA at the TIEMPO research center (St. Larrea, Buenos Aires, Argentina). The equipment was a Lunar Prodigy Avance. Most of the studies on bone properties (dielectric, mechanical, ultrasonic, etc.) are correlated with BV/TV. In this paper we obtain the BV/TV parameter by using the relation obtained by Tassani et al. \cite{tassani2011volume}. They found that the average value of the tissue mineral density (TMD) is approximately constant with respect to the BV/TV. It was around 1.22 g / cm${^3}$ for samples of donors with mean age of 74 years old (similar to the samples of this paper). Using the relation:
\begin{equation}
TMD = \frac{BMC}{\text{Bone Volume}} \approx 1.22 \text{g / cm}^3
\end{equation}
we can obtain the bone volume (BV) and then make the ratio measuring the total volume (TV) of the samples with a common caliber.
\subsection{Dielectric measurements}
\label{DP}
\begin{figure}
\centering
\includegraphics[width=0.8\textwidth]{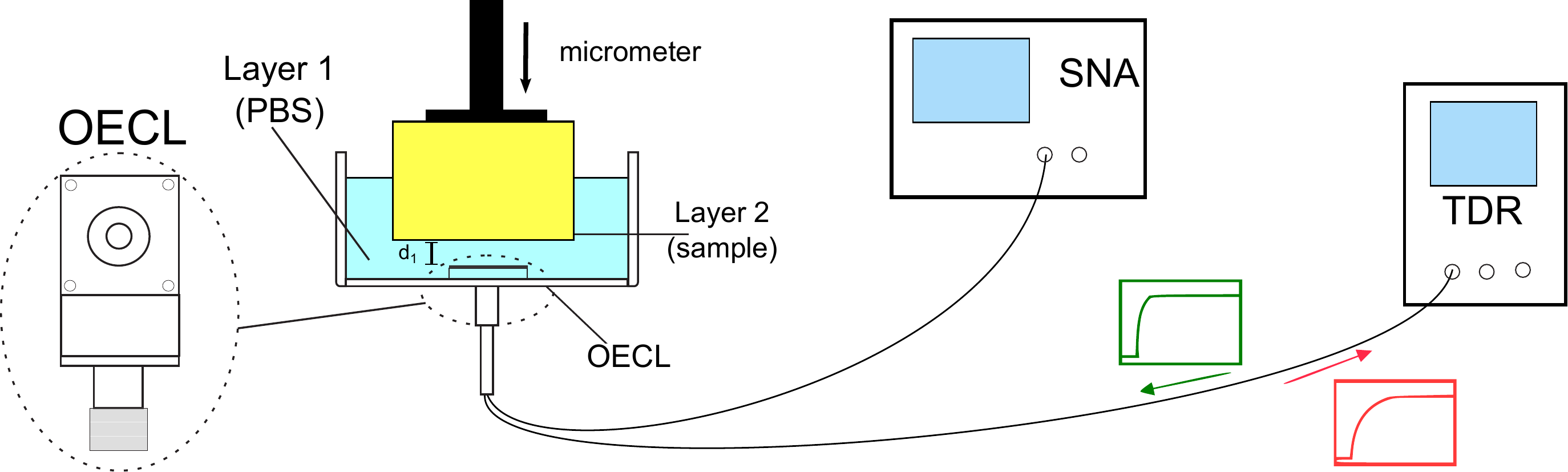}
\caption{Measurement setup. From left to right: the OECL, the sample cell with the first layer (PBS) and second layer (trabecular sample), the Scalar Network Analyser and the Time Domain Reflectometer.}
\label{Fig1}
\end{figure}

\begin{figure}
\centering
\includegraphics[width=1.0\textwidth]{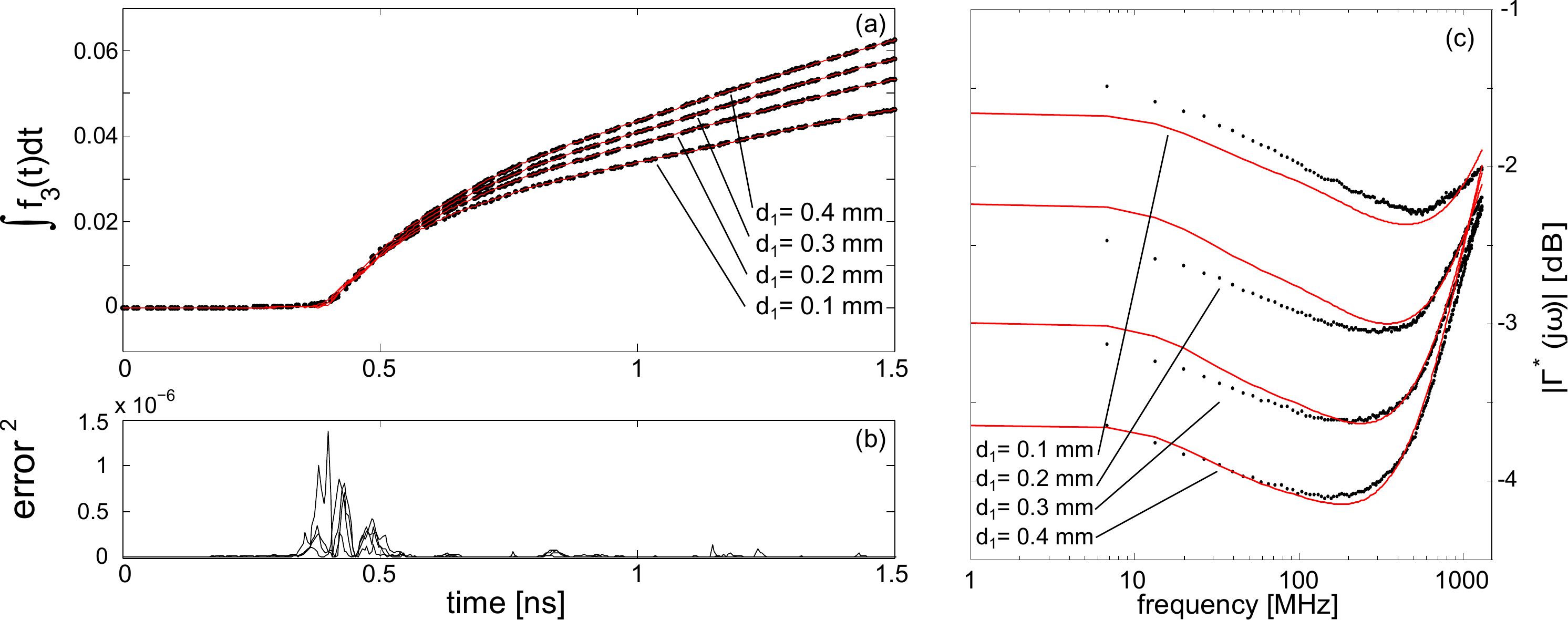}
\caption{Measurement procedure. (a) Output measured (dashed-dotted line) and estimated by system identification technique (continuous line) for several thicknesses of the first layer. (b) Error calculated by subtracting the signals shown in (a). (c) Validation data. The reflected magnitude measured with SNA (dotted line) andthe estimated by the identified model (continuous line).}
\label{Fig2}
\end{figure}

\begin{figure}
\centering
\includegraphics[width=0.7\textwidth]{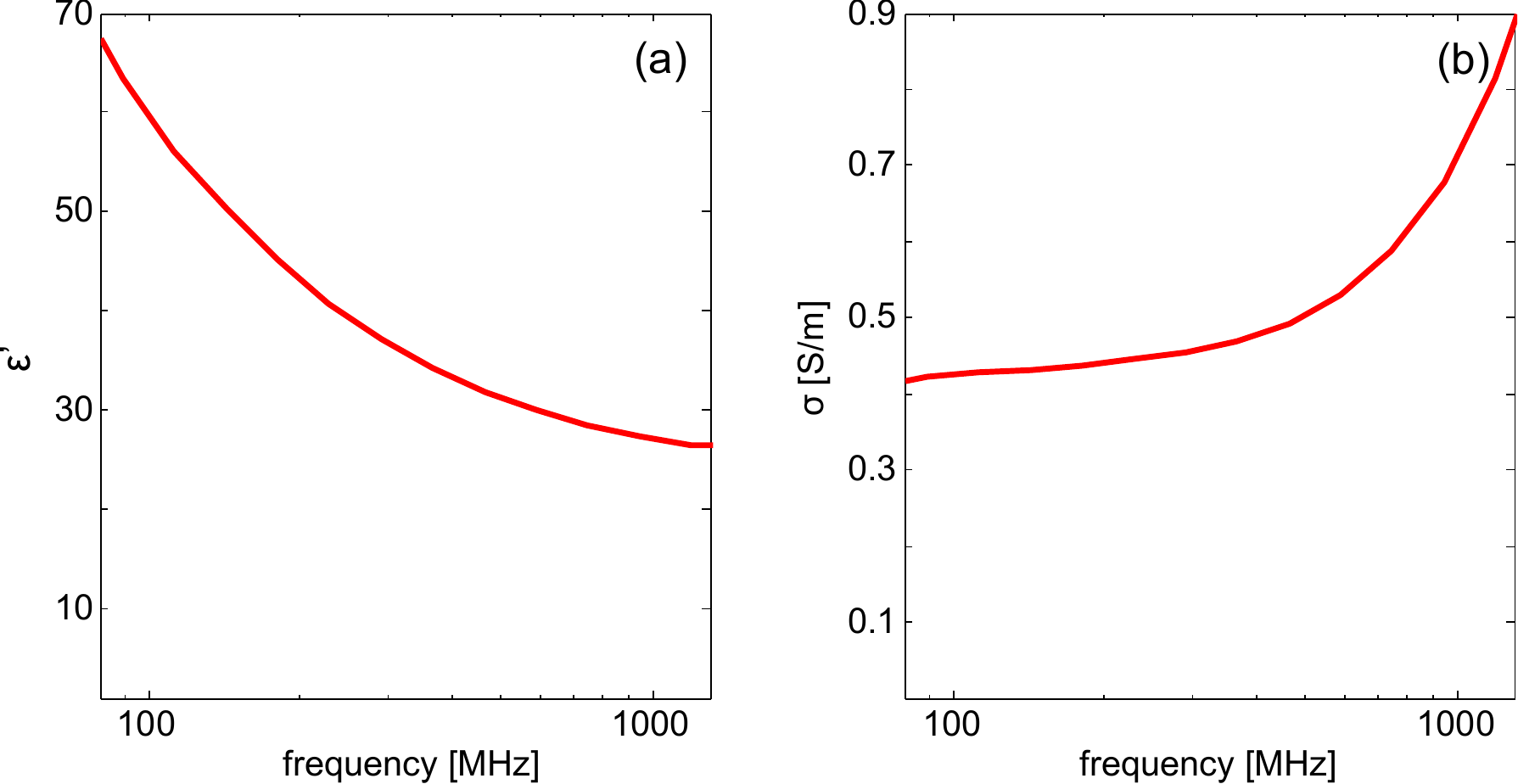}
\caption{Mean value of the dielectric properties of the measurement presented in Fig.\ref{Fig2}. (a) Relative permittivity and (b) conductivity in function of frequency.}
\label{PERMIT}
\end{figure}

OECL have been extensively used in characterization of biologic tissues \cite{hagl2003sensing}, actually most of the measurements of \cite{gabriel1996a,gabriel1996b,gabriel1996c} were performed with this probe. In this paper, OECL together with time domain methods are used here to obtain the dielectric properties of samples. Figure \ref{Fig1} shows the setup. Time-Domain (TD) methods are based on the measurement of signal responses as a function of time from a specimen when it is excited with a transient signal. The technique used in this work combines a differential method with system identification and was presented in \cite{irastorza2011,ivancich1992}. Briefly,
the equation which relates the apparent complex permittivity of the sample ($\varepsilon_{app}(s)$) and the Laplace transform of the measured signals can be written as:
\begin{equation}
 H(s)= \dfrac{Y(s)}{U(s)} = Z_{0}\cdot\left[C_{0}\varepsilon_{app}(s)+C_{f}\right]+\dfrac{f_{3}(\infty)}{s\cdot f_{2}(\infty)} = \dfrac{F_{3}(s)}{s\cdot F_{2}(s)}
 \label{eqn1}
\end{equation}
where $s$ is the Laplace variable. $F_{2}(s)$ and $F_{3}(s)$ are the transforms of the difference between sample ($R(t)$) and short circuit (that is, minus the applied voltage; $-V_{0}(t)$) and the difference between the open circuit ($V_{0}(t)$) and sample, respectively. $C_{0}$, $C_{f}$, and $Z_{0}$ are the capacitance of the probe, fringe capacitance, and the characteristic impedance of the line, respectively. In Eq.\ref{eqn1} we defined $Y(s)=s^{-1} F_{3}(s)$ and $U(s)=F_{2}(s)$ as the output and input of a dynamic system. When $H(s)$ is approximated by a rational transfer function it takes the form:
\begin{equation}
 H(s)\approx \dfrac{B(s)}{F(s)} = \dfrac{b_{0}s^{m}+b_{1}s^{m-1}+\cdots + b_{m}}{s^{n}+a_{1}s^{n-1}+\cdots + a_{n}}
 \label{eqn2}
\end{equation}
where $B(s)$ and $F(s)$ are polynomials of $s$ and $m$ and $n$ can be both integer or not. The latter represents fractional systems which have demonstrated good results for dielectric measurements on biological tissues \cite{irastorza2011}. $a_{i}$ and $b_{i}$ are real constants. Using OECL and measuring a bilayer structure the apparent permittivity can be approximated by (see \cite{irastorza2009noninvasive}):
\begin{equation}
\varepsilon_{app}(s) \approx \left[\varepsilon_{1}(s)-\varepsilon_{2}(s)\right] \cdot \left[1-e^{-q\cdot d_{1}}\right] + \varepsilon_{2}(s) \text{  when  } \varepsilon_{1} > \varepsilon_{2}
 \label{eqn3}
\end{equation}

\begin{equation}
\frac{1}{\varepsilon_{app}(s)} \approx \left[\frac{1}{\varepsilon_{1}(s)}-\frac{1}{\varepsilon_{2}(s)}\right] \cdot \left[1-e^{-q\cdot d_{1}}\right] + \frac{1}{\varepsilon_{2}(s)} \text{  when  } \varepsilon_{1} < \varepsilon_{2}
 \label{eqn4}
\end{equation}

where $\varepsilon_{1}(s)$, $d_{1}$, and $\varepsilon_{2}(s)$  are the permittivity and thickness of the first layer, and the permittivity of the second layer, respectively. The constant $q$ is empirical and depends on the probe size (in this paper $q = $ 1.5192 mm$^{-1}$, see the reference \cite{irastorza2009noninvasive}). The usefulness of Eqs. \ref{eqn3} and \ref{eqn4} can be shown as follows. Suppose that we have to measure a sample of $\varepsilon_{2} = $ 40 with an OECL. If a small gap of air ($\varepsilon_{1} = $ 1) of thickness $d_{1}=$ 0.1 mm is between the probe and the sample, then using Eq.\ref{eqn4}, the apparent permittivity is $\varepsilon_{app} \approx  $ 6.15. If instead of air there is water ($\varepsilon_{1} =$ 78), Eq. \ref{eqn3} gives  $\varepsilon_{app} \approx $ 45.35. Now we suppose an error $\Delta d$ on the measurement of the thickness $d_{1}$, which may represent an air gap with unknown thickness (bad contact of the probe). With $\Delta d \approx$ 0.02 mm the 
obtained errors are $\approx$ \%20 or $\approx$ \%2, when air or water is present, respectively. Consequently, we consider that measuring bone with a bilayer structure using as the first layer the PBS solution (with known dielectric properties) significantly reduces the contact error between the probe and the sample.\\
The measurement procedure can be summarized as follows:
\begin{description}
\item[i.] obtain the probe parameters: $C_{0}$, $C_{f}$ and $q$ with alcohols and bilayer structures (alcohols and teflon plates) with known dielectric properties (not shown in this paper, see \cite{irastorza2009noninvasive}). This calibration by immersing in standard liquids is widely used with OECL \cite{gabriel2006dielectric}.
\item[ii.] measure a bilayer structure (PBS / bone sample) with different thicknesses of the first layer (from $d_{1}$= 0.1 to 0.6 mm with step of 0.1 mm), hence six measurements of the apparent permittivity of each sample are obtained.
\item[iii.]  measure the PBS layer dielectric properties by moving away the sample (far enough $d_{1} > 7$ mm, see \cite{hagl2003sensing}).
\item[iv.]estimate the dielectric properties of the sample by using Eq. \ref{eqn3}. Note that we finally obtain six measurement of each sample then we calculate mean and standard deviation of the dielectric parameters.
\end{description}
Four measurements are shown in Fig. \ref{Fig2} (a)-(b) with their respective validation in Fig.\ref{Fig2} (c). Figure \ref{PERMIT} shows the mean value of the dielectric properties in function of frequency obtained from the data of Fig. \ref{Fig2}.\\
In time domain measurements a Hewlett-Packard TDR/Sampler 1815B plug in a Hewlett-Packard 1801A oscilloscope and a sampler head HP1106A with a tunnel diode HP1106 was used (identification data). The reflection coefficient magnitude was also measured in frequency domain by the Scalar Network Analyzer (SNA) HP8711A (Hewlett-Packard) up to 1300 MHz (see Fig.\ref{Fig1}). An APC7 Hewlett-Packard coaxial was used connected to a custom-built OECL with 7 mm output diameter (see Fig.\ref{Fig1}). The calibration gives values $C_{0} = 0.0511$ pF and $C_{f} = 0.036$ pF (see \cite{irastorza2011} for details on calibration).\\
The experiments were performed at 25$^{o}$C and temperature was controlled by a cryostat LAUDA RE106 (Lauda-K\"{o}nigshofen, Germany). It should be remarked that the temperature coefficients for permittivity and conductivity are tissue-type and frequency dependent. For example: the dielectric properties of liver have been shown to vary linearly with temperature at 915 MHz (see \cite{lazebnik2006ultrawideband}), with coefficients $\approx$ -0.1 \% $^{o}$C$^{-1}$ and 1.33 \% $^{o}$C$^{-1}$, for permittivity and conductivity, respectively. For trabecular bone, the information on these coefficients is scarce and not sufficient in literature, but we assume that the differences between 25 $^{o}$C and 37 $^{o}$C are small enough.

\subsection{Simulation}
\label{SIM}
\begin{figure}
\centering
\includegraphics[width=0.8\textwidth]{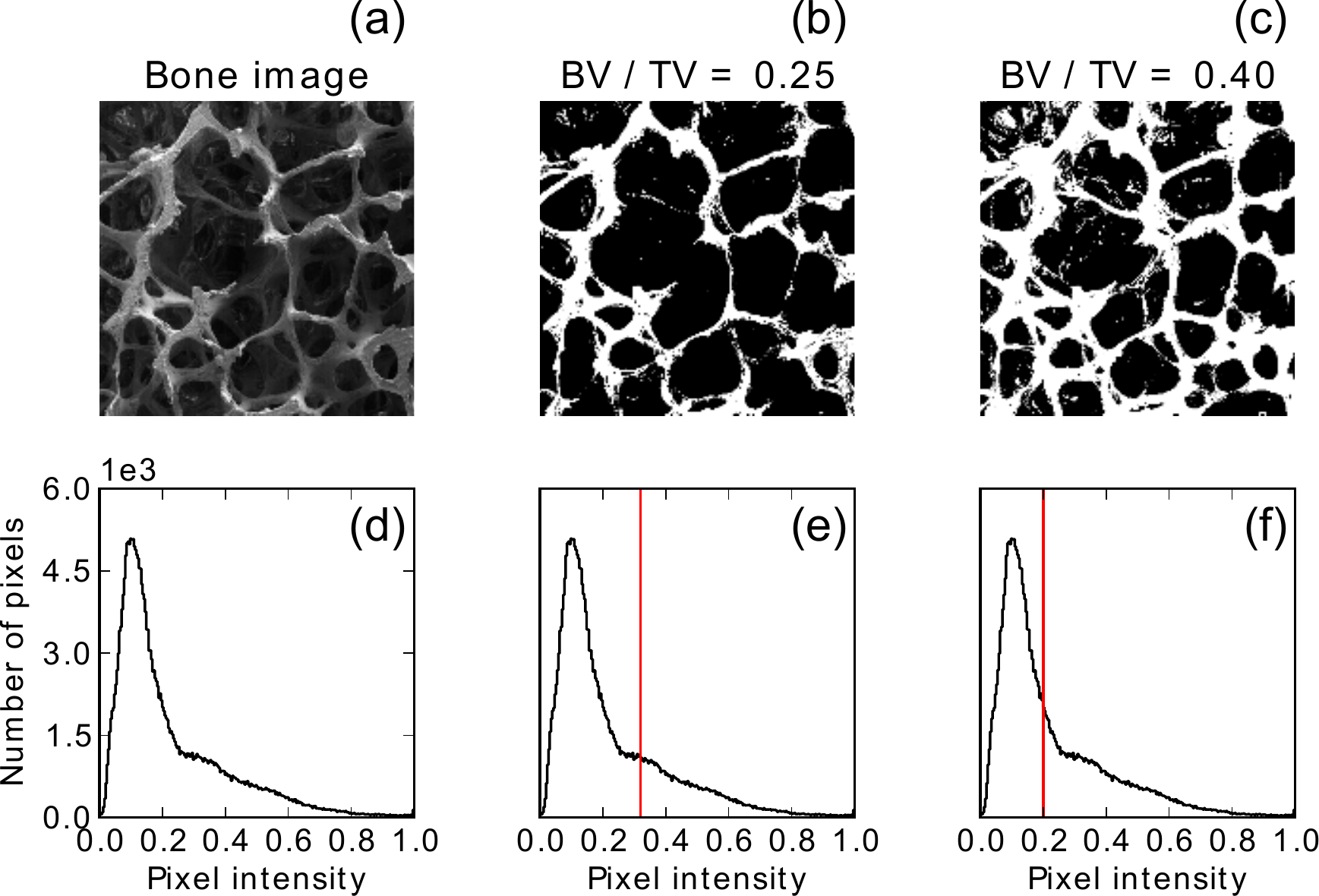}
\caption{Simulation samples. (a) SEM image of bone sample and (d) is its histogram. (b) and (c) are examples of the simulated media with BV/TV = 0.25 and 0.4 respectively. In (e) and (f) are marked with a vertical continuous line the threshold for the media of figure (b) and (c), respectively.}
\label{FIGSIMU}
\end{figure}

The effective dielectric properties of a single sample were simulated. They were performed using a realistic model from a scanning electron microscopy (SEM) image of an elderly bone (osteoporotic woman 80 years old). The image was obtained from \cite{hansmalab} and it was already used by \cite{Golden2011} for simulation purposes (see Fig. \ref{FIGSIMU} (a)). This image was selected for two reasons. First to check the validity of the simulations by comparison to the results already published by \cite{Golden2011}. Second, it should be remarked that the dielectric measurement technique (using OECL) can not detect the anisotropy, thereafter the selected sample is isotropic (the degree of anisotropy is approximately 1, see \cite{irastorza2013}). The simulation procedure was previously described and validated \cite{irastorza2013}. Briefly, effective dielectric properties calculation in 2D was studied by observing the reflection of a slice of the realistic model of trabecular bone. In order to obtain the effective dielectric properties, the reflection coefficient of the realistic bone model was compared with the reflection coefficient of a homogeneous slice (with the effective dielectric properties) in the frequency range of interest. Iterations were conducted until the same reflection coefficient was reached. Maxwell equations are numerically solved using finite-difference time-domain (FDTD) method \cite{TafloveBook}.  It was implemented in the software package MEEP (Massachusetts Institute of Technology, Cambridge, USA \cite{meep}). The tissue was modelled as a two components medium: the PBS and the trabecular solid bone matrix. For the dielectric parameters of PBS, the Debye model was estimated from the experimental measurements. The dielectric properties of trabeculae were obtained from references \cite{gabriel1996a,gabriel1996b,gabriel1996c}. In order to simulate samples with different BV/TV values, the threshold of the grey scale histogram used to define the two components medium was changed (see Fig. \ref{FIGSIMU} (e-f)) \cite{scikits}. In this way we cover BV/TV values from 0.2 to 0.4. 

\subsection{Data analysis}
Statistical analyses were conducted with custom built programs in Python 2.7 Release (http://www.python.org/). The linear correlation coefficients between BV/TV and relative permittivity, and between BV/TV and conductivity were determined by using the Pearson correlation analysis.
\section{Results}
\label{RESUL}
\begin{figure}
\centering
\includegraphics[width=1.0\textwidth]{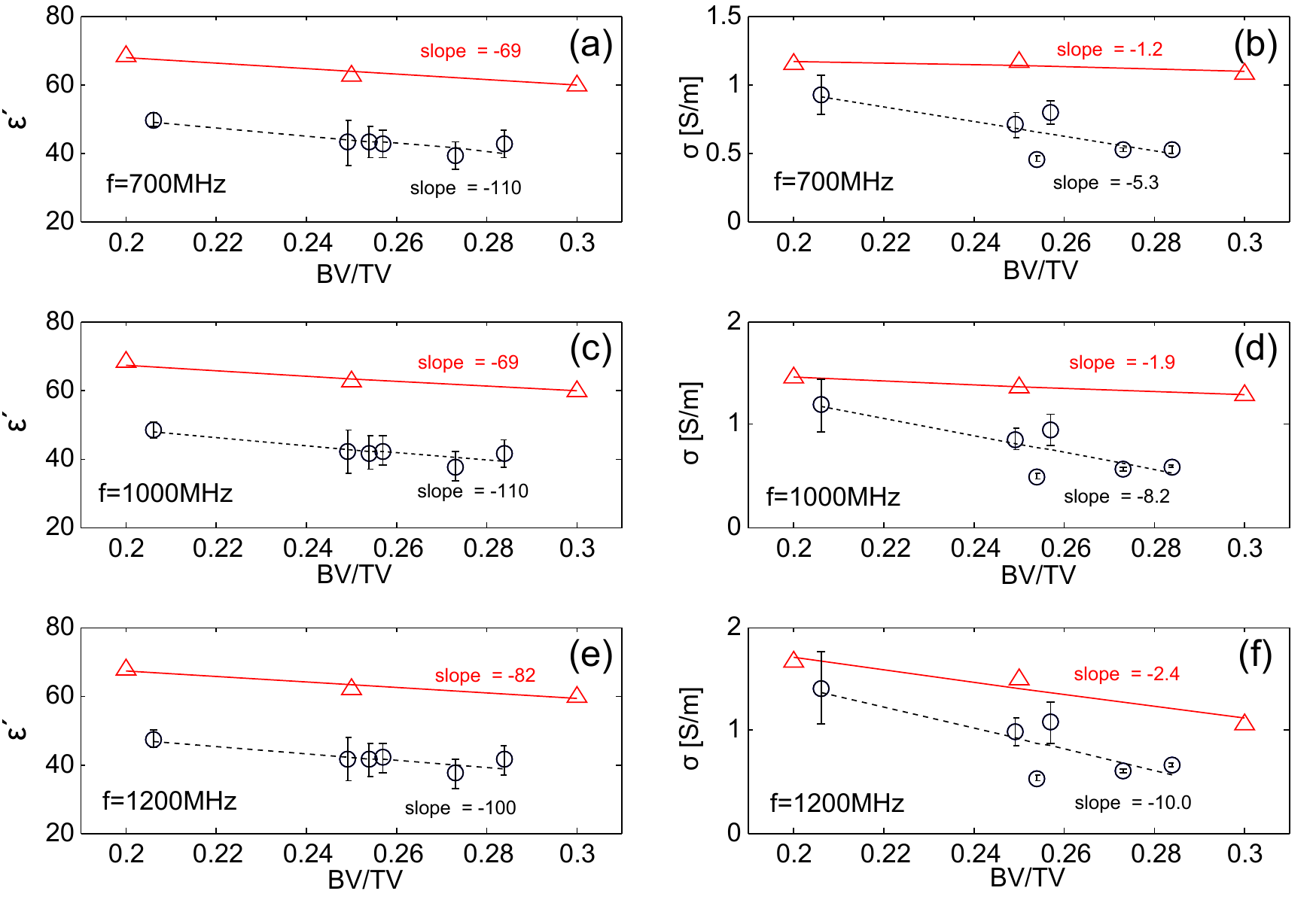}
\caption{Relation between BV/TV parameter and dielectric properties for selected frequencies. In triangle are plotted the simulation data and in circles the measured one (error bar corresponds to standard deviation). Continuous and dashed lines show the linear estimates for the simulated and measured data, respectively. The slope of the simulation data is calculated with the BV/TV from 0.2 to 0.4.}
\label{Fig6}
\end{figure}

Table \ref{Table I} shows the mean values (standard deviation) of the measured values for some selected frequencies. \\
\begin{table}
\centering
\caption{Mean values (SD) for dielectric properties at fixed frequencies. BMD (SD) = 0.268 (0.028) g / cm$^2$ and BV/TV (SD) = 0.254 (0.027).}
\label{Table I}
\begin{tabular}{ccc} 
Frequency (MHz) & Relative permittivity & Conductivity (Sm$^{-1}$) \\
\hline
700 & 43.29 (3.39) & 0.653 (0.181)\\
1000 & 42.22 (3.40) & 0.771 (0.270)\\
1200 & 41.72 (3.24) & 0.870 (0.339)
\end{tabular}
\end{table}
As was detailed in section \ref{DP}, the dielectric parameters of each sample were measured using six thicknesses of the first layer. Each measurement yielded an estimate from time domain, and then it was validated in frequency domain. After the six measurement were obtained, the mean values and the standard deviation were calculated and finally plotted in Figure \ref{Fig6}. The range of the BV/TV measured data goes from 0.206 to 0.284. Figure \ref{Fig6} also shows the simulated values. In the figure are only shown the simulated data near the samples BV/TV range. The linear estimates are plotted as well. For example, the slopes of the plot of BV/TV versus relative permittivity at 1200 MHz are approximately -100 and -80 for the measured and simulated values, respectively. In the case of the BV/TV versus conductivity at the same frequency, the slopes are around -10 Sm$^{-1}$ and -2.4 Sm$^{-1}$ for the measured and simulated, respectively.\\
The whole simulated data are shown in Table \ref{Table II}. The BV/TV range was simulated from 0.2 to 0.4 with a step of 0.05. The selected frequencies are the ones selected for the measured data. The degree of anisotropy of the sample was obtained previously with the mean intercept length criteria \cite{irastorza2013}. It resulted in an almost isotropic sample. Such condition has been thought on purpose due to the experimental limitation.\\
\begin{table}
\centering
\caption{PBS-bone tissue simulation results with an elderly trabecular bone sample. The degree of anisotropy is 1.08.}
\label{Table II}
\begin{tabular}{cccc} 
Frequency (MHz) & Relative permittivity & Conductivity (Sm$^{-1}$)& BV/TV \\
\hline
\multirow{4}{*}{700} & 68.17 & 1.150 & 0.20\\
    & 62.56 & 1.167 & 0.25\\
    & 59.98 & 1.074 & 0.30\\
    & 56.64 & 1.025 & 0.35\\
    & 53.77 & 0.933 & 0.40\\
\hline
\multirow{4}{*}{1000} & 67.33 & 1.448 & 0.20\\
    & 63.03 & 1.370 & 0.25\\
    & 59.64 & 1.282 & 0.30\\
    & 55.88 & 1.187 & 0.35\\
    & 51.36 & 1.068 & 0.40\\
\hline
\multirow{4}{*}{1200} & 67.66 & 1.659 & 0.20\\
    & 61.86 & 1.497 & 0.25\\
    & 59.55 & 1.059 & 0.30\\
    & 52.62 & 1.235 & 0.35\\
    & 51.67 & 1.182 & 0.40
\end{tabular}
\end{table}
Table \ref{Table IV} shows the statistical results for the linear regression analysis. The sample Pearson correlation coefficients ($r$) and the coefficients of determination ($r^{2}$) were computed. The statistical analysis was performed for experimental and simulation data.
\begin{table}
\centering
\caption{Linear correlation analysis. Coefficient of determination ($r^{2}$) between the dielectric properties and BV/TV of human trabecular bone samples. The p-values were calculated. The values shown without mark are not significant.}
\label{Table IV}
\begin{tabular}{lllll} 
\multirow{2}{*}{Frequency (MHz)} &\multicolumn{2}{c}{Relative Permittivity}&\multicolumn{2}{c}{Conductivity}\\
 & Experiment & Simulation & Experiment & Simulation \\
 \hline
700 & 0.77* & 0.97** & 0.60 & 0.90* \\
1000 & 0.75* & 0.99** & 0.65* & 0.99** \\
1200 & 0.72* & 0.95** & 0.67* & 0.61 \\
\hline
\footnotesize *p $<$ 0.05\\ **p $<$ 0.01 
\end{tabular}
\end{table}

\section{Discussion}
\label{DISCUSS}
The main assumption in this paper is that there is not a significant difference between trabecular and cortical adult bone tissues in terms of tissue mineral density, which remains constant for a broad range of BV/TV \cite{tassani2011}. This limitation implies that the tissue must be homogeneous, and as a consequence the BV/TV parameter can be calculated from BMD measurements. The bone volume fractions obtained in the samples are closely in agreement with published data \cite{sierpowska2006,tassani2011,hildebrand1999}.\\
\begin{table}
\centering
\caption{Comparison with literature values.}
\label{Table III}
\begin{tabular}{cccccc} 
Frequency & Relative & Conductivity &  \multirow{2}{*}{BV/TV}  & \multirow{2}{*}{Reference}\\
(MHz) & permittivity & (Sm$^{-1}$)& & \\
\\
\hline
\multirow{2}{*}{950} & 7.10-7.50 & 0.070-0.077 & 0.25 & \multirow{2}{*}{simulation \cite{irastorza2013}}\\
 & 10.34-11.44 & 0.149-0.154 & 0.54 & \\
\hline
1000 & $\approx$ 7.2 & $\approx$ 0.06 & 0.08-0.11 & simulation \cite{bonifasi2009} \\
\hline
900 & 60-40 & 1.8-1.3 & 0-0.45 & \multirow{3}{*}{in vitro porcine\cite{meaney2012bone}} \\
1100 & 60-31 & 2.5-1.9 & 0-0.45 &  \\
1300 & 55-30 & 2.7-1.9 & 0-0.45 &  \\
\hline
\multirow{2}{*}{1000} & 20.584 & 0.363 & - & in vitro ovine \cite{gabriel1996a,gabriel1996b,gabriel1996c}\\
 & 5.485 & 0.043 & - &  bone marrow \cite{gabriel1996a,gabriel1996b,gabriel1996c}\\
\hline
 \multirow{4}{*}{1300} & 12.5 & 0.53 & - & \multirow{4}{*}{in vivo\cite{meaney2012}}\\
 & 13.6 & 0.84 & - & \\
  & 13.5 & 0.80 & - & \\
  & 16.7 & 0.92 & - &  
\end{tabular}
\end{table}
The values of dielectric properties of human trabecular bone at microwave frequencies have a relative controversy. There exist a reduced number of papers that treat this topic. Table \ref{Table III} shows the experimental and simulation data collected from literature. The classical values published by Gabriel et al. \cite{gabriel1996a,gabriel1996b,gabriel1996c} are 20.58 and 0.36 Sm$^{-1}$ for 1000 MHz. They were obtained from ovine bone. Sierpowska \cite{sierpowska2007thesis} measured human trabecular bone in this frequency range (not shown in the table) but the results remains unpublished. The obtained values are approximately 15 and 0.350 Sm$^{-1}$ (extracted from figures). It was concluded that the superficial surface layer contains more damage and liquid than the bulk sample, therefore it may lead to higher values of permittivity and conductivity. The permittivity and conductivity mean values measured in this work for 1000 MHz are around 42 and 0.8 Sm$^{-1}$, respectively. These values are still higher 
than reference \cite{sierpowska2007thesis}, but they are close enough from those obtained by Meaney et al. \cite{meaney2012bone} in porcine samples embedded in 0.9\% saline solution. They obtained (in the BV/TV range of this paper) values between 30 and 45, and between 1.9 Sm$^{-1}$ and 2.3 Sm$^{-1}$, for the permittivity and conductivity, respectively. Remarkably, the experimental set-up of reference \cite{meaney2012bone} is quite different to the one used in this work. The \textit{in vivo} measurements of the permittivity \cite{meaney2012} are in the range of 12 and 16, near to the measured in \cite{sierpowska2007thesis}. For the \textit{in vivo} conductivity values of Meaney et al. \cite{meaney2012}, they are close to the measurements of this work.\\
The trabecular bone tissue in the reconstruction algorithms of microwave tomography is usually taken as an isotropic effective medium. On the other hand, in the reference \cite{bonifasi2009} the authors simulated the human trabecular tissue from micro tomography images as a two components medium. They obtained much lower values than those measured \textit{in vivo}. Similar results were obtained in the simulation work of our group \cite{irastorza2013}. This difference with \textit{in vivo} values is mainly due to the dielectric parameters used for each phase \cite{gabriel1996a,gabriel1996b,gabriel1996c}. The simulation values obtained in this work ($\varepsilon ' \approx 63$ and $\sigma \approx 1.4$ Sm$^{-1}$) are higher than \textit{in vivo} values. Nevertheless, if they are compared to those measured for porcine samples in similar condition (saline solution plus trabecular matrix) the values are approximate.\\
A special commentary should be addressed about the linear correlation between the BV/TV versus permittivity and conductivity. Negative high correlations were found between BV/TV and relative permittivity at 700, 1000 and 1200 MHz for the experimental data ($p < 0.05$ and $r^{2}>0.7$) which is in agreement with the simulated data (see the details in Fig. \ref{Fig6} and Table \ref{Table IV}). The linear negative correlation of the BV/TV versus conductivity is not as good as for the permittivity. The best values occur at 1200 MHz with $r^{2}=0.67$ and $p < 0.05$ (see Fig. \ref{Fig6} (f)). However, regarding this slope, the simulations presented in this work are in close agreement to reference \cite{meaney2012}.\\
As concluding remarks, we can stand out two important issues: first regarding the dielectric model for simulation purposes, and second, regarding the linear negative correlation of the BV/TV versus permittivity (experiments and simulations). For the former, we can say that if the trabecular tissue is simulated as a two components medium then the dielectric parameters of the constituent have to be reformulated. The conductivities values of the references \cite{bonifasi2009,irastorza2013} are  much lower than the \textit{in vitro} and \textit{in vivo} measurements. The simulations of this work yield closer values for the conductivity. This is not the case for the permittivity, whose values are higher than the \textit{in vivo} data. However, they are comparable to the \textit{in vitro} measurements. In the second issue, we addressed the relationship between BV/TV and permittivity (conductivity) with experiments and simulations, and both approaches show evidence of a negative relation: the higher the BV/TV the 
lower the permittivity (conductivity).

\section*{Acknowledgments}
This work was supported by grants from Universidad Nacional de La Plata, Consejo Nacional de Investigaciones Cient\'ificas y
T\'ecnicas and Agencia Nacional de Promoci\'on Cient\'ifica y Tecnol\'ogica of Argentina. C.M.C. and F.V. are members of CONICET. We would like to thank the collaboration of Dr. Sergio Valente from Gabinete de Biomec\'{a}nica, Dto. de Ing. Mec\'{a}nica, Facultad de Ingenier\'{i}a, Universidad of Buenos Aires, Argentina. This paper is in memoriam of Dr. Zulema Man from the TIEMPO research center, Buenos Aires, Argentina.

\section*{Conflict of interest statement}
No conflict of interest.

\section*{Ethical approval}
Not required.


\end{document}